\documentclass{emulateapj}

\usepackage{graphicx}

\usepackage{times}



\def\keV{ke\kern-0.05emV}
\newcommand{\chandra}{\emph{Chandra}}

\renewcommand{\arcsec}{\ensuremath{''}}

\def\chandra    {{\em Chandra}\/}
\def\rxj1720    {{ RXJ1720.1+2638}\/}

\def\degd       {$^{\circ}\!$}
\def\second     {{\prime\prime}}
\def\2A     {{2A 0335+096}}
\def\ms1455     {{scemo}}

\begin{document}

\submitted{Submitted to ApJ}

\title{A \chandra ~ Study of the Complex  Structure in the Core of 2A 0335+096}

\author{P.\ Mazzotta\altaffilmark{1,2}, A.\ Edge\altaffilmark{1}, 
M.\ Markevitch\altaffilmark{2}}
\altaffiltext{1}{Department of Physics,
University of Durham, South Road, Durham DH1 3LE; pasquale.mazzotta@durham.ac.uk}
\altaffiltext{2}{Harvard-Smithsonian Center for Astrophysics, 60 Garden St.,
Cambridge, MA 02138}

\shorttitle{\chandra ~ Study  of 2A 0335+096}
\shortauthors{MAZZOTTA ET AL.}

\begin{abstract}

We present a \chandra ~ observation of the central ($r< 200$~kpc) region of
the  cluster of galaxies \2A , rich in interesting phenomena.  On large
scales ($r> 40$~kpc), the X-ray surface brightness is symmetric and slightly
elliptical.  The cluster has a cool dense core; the radial temperature
gradient varies with position angle.  The radial metallicity profile shows a
pronounced central drop and an off-center peak.  Similarly to many clusters
with dense cores, \2A ~ hosts a cold front at $r\approx 40$~kpc south of the
center.  The gas pressure across the front is discontinuous by a factor
$A_P=1.6\pm0.3$, indicating that the cool core is moving with respect to the
ambient gas with a Mach number $M\approx 0.75\pm 0.2$.  The central dense
region inside the cold front shows an unusual X-ray morphology, which
consists of a number of X-ray blobs and/or filaments on scales 
$\gtrsim 3$~kpc, along with two prominent X-ray cavities.  The X-ray
blobs are not correlated with either the optical line emission
(H$\alpha+$[N\textsc{ii}]), member galaxies or radio emission.  Deprojected
temperature of the dense blobs is consistent with that of the less dense
ambient gas, so these gas phases do not appear to be in thermal pressure
equilibrium.  An interesting possibility is a significant, unseen non-thermal
pressure component in the inter-blob gas, possibly arising from the activity
of the central AGN.  We discuss two models for the origin of the gas blobs
--- hydrodynamic instabilities caused by the observed motion of
the gas core, or ``bubbling'' of the core caused by multiple outbursts of the
central AGN.

\end{abstract}


\section{Introduction}

The cores of clusters contain great dynamical complexity. The presence of
gas cooling (Fabian 1994), relativistic plasma ejection from the central
galaxy  (McNamara et al.  2000,  Fabian et al. 2000), 
the action of magnetic fields (Taylor, Fabian \& Allen, 2002), possible
thermal conduction (Voigt et al. 2002), local star-formation (Allen et al.
1995) and substantial masses of cold ($T<30$~K) molecular gas (Edge 2001)
all complicate the simple hydrostatic picture. 

The launch of \chandra ~ and {\it XMM-Newton} have allowed the cores of
clusters to be studied in unprecedented detail spatially and
spectrally. These advances have led to a number of important results (e.g.,
Peterson et al.\ 2001; Forman et al. 2002, McNamara 2002). The very
restrictive limits on low-temperature gas from {\it XMM-Newton} RGS
spectra are forcing us to re-examine some of the basics paradigms of the
cooling flows (e.g. Kaastra et al. 2001; Peterson et al. 2001, Tamura et
al. 2001, Molendi \& Pizzolato 2001).
The spatial resolution of  \chandra ~ can be used
to study the brightest, nearby cooling flows to advance greatly our
understanding of the processes occurring in these regions.
In this paper we present the \chandra ~ observation of the  
Zwicky cluster  Zw~0335.1+0956 whose properties may help us to address 
some of the above issues

The  Zw~0335.1+0956 was first detected as a strong
X-ray source by {\it Ariel-V} (Cooke et al. 1978) and we therefore use the
traditional identification of 2A~0335+096 in this paper.  2A~0335+096 is
among the brightest 25 clusters in the X-ray sky (Edge et al. 1990).
The presence of a cooling flow was noted by Schwartz, Schwarz, \& Tucker (1980)
 and its X-ray properties have been studied extensively 
over the past two decades
(Singh, Westergaard \& Schnopper 1986, 1988; White et al. 1991; Sarazin
O'Connell \& McNamara 1992; Irwin \& Sarazin 1995, Kikuchi et al. 1999). 
The optical properties of the central galaxy in 2A~0335+096 have
been studied in detail (Romanishin \& Hintzen 1988; 
McNamara, O'Connell \& Bregman 1990) and the strong, 
extended optical line emission marks this system out
as an atypical elliptical galaxy  but a prototypical 
central galaxy in a cooling flow. A deep, multi-frequency radio study of
2A~0335+096 (Sarazin, Baum \& O'Dea 1995) shows a weak, flat-spectrum radio
source coincident with the dominant galaxy which is surrounded by an
amorphous steep-spectrum `mini-halo'. The tentative detection of HI
absorption (McNamara, Bregman \& O'Connell 1990) and firm detection of CO
emission (implying 2$\times 109$ M$_\odot$ of molecular gas) 
and IRAS 60~$\mu$m continuum (Edge 2001) further highlight this
cluster as one for detailed study. The implied mass deposition
rate from the CO detection is low ($<$5  M$_\odot$~yr$^{-1}$)
if the cold molecular gas found is deposited in the cooling flow.

We use $H_0=70$~km~s$^{-1}$~kpc$^{-1}$, $\Omega=0.3$, and $\Lambda=0.7$,
which imply a linear scale of
0.7~kpc per arcsec at the distance of 2A 0335+096 ($z=0.0349$).
Unless specified otherwise, all the errors are at $90\%$ confidence level
 for one interesting parameter.

\begin{figure*}
\centerline{\includegraphics[width=0.95\linewidth]{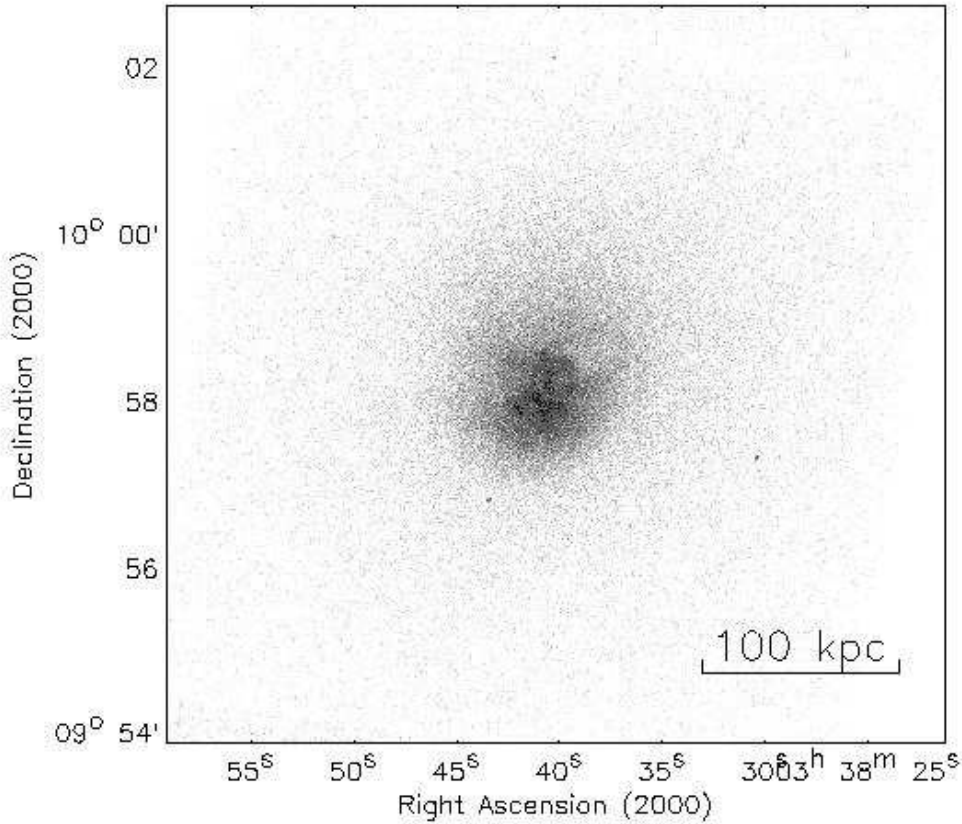}}
\caption{Photon image of the central region (ACIS-S3) of \2A ~ in the
0.3-9~keV energy  band. The image is  background subtracted, vignetting 
corrected, and  binned to $1\arcsec$ pixels. 
The image shows wealth of substructures in the innermost 50~kpc region.
Moreover, it shows two significant surface brightness depressions (cavities) 
on opposite sides of  the X-ray peak in the direction North-West
South-East. We also notice a surface brightness edge to the South.}
\label{fig:x-rayimage}
\end{figure*}

\section{Data preparation}\label{par:data} 

\2A ~ was observed  on 06 Sept 2000 
with the Advanced CCD Imaging Spectrometer (ACIS) 
using the  back-illuminated chip S3.
The total exposure time is $21.4$~ksec.
In this paper we
concentrate on the bright central $r\lesssim 200$~kpc region of the cluster
which lies fully within the ACIS-S3 chip.

Hot pixels, bad columns, chip node boundaries, and events with grades 1, 5,
and 7 were excluded from the analysis.  We cleaned the observation of
background flares following the prescription given in
Markevitch et al.\ (2003).
Because chip S3 contains the bright cluster
region, we extracted a light curve for the other backside-illuminated chip
S1 in the 2.5--6 keV band, which showed that most of the observation is
affected by a mild flare. Because the region of the cluster under study is
very bright, we chose a less conservative flare cleaning than is
recommended, excluding the time intervals with rates above a factor of 2 of
the quiescent rate (taken from the corresponding blank-sky background
dataset, Markevitch 2001\footnote{http://asc.harvard.edu/ 
``Instruments and Calibration'', ``ACIS'', ``ACIS Background''}) 
instead of the nominal factor 1.2. This resulted
in the final exposure of 13.8 ks.  During the accepted exposure, the
frontside-illuminated chips did not exhibit any background rate increases, 
therefore the residual background flare is of the ``soft'' species
and can be approximately modeled as described by Markevitch et al.\
(2003). We fitted the excess background in chip S1 above 2.5 keV with the
flare model and included it, rescaled by the respective solid angle, to the
cluster fits.  The main component of the background was modeled using the
above-mentioned quiescent blank-sky dataset, normalized by the rate in the
10--12 keV band (which can be done because the residual flare in our
observation is ``soft'' and does not affect the high-energy rate).  The
addition of the flare component has a very small effect on the results,
because the cluster core is so bright.  We therefore ignored the possible
(not yet adequately studied) spatial nonuniformity of the flare component.

\section{X-ray Imaging Analysis}\label{par:image}

Fig.~\ref{fig:x-rayimage} shows a background subtracted, vignetting
corrected image of the central region of the cluster, extracted in the
0.3-9~keV energy band and binned to $1$\arcsec ~ pixels.
The cluster X-ray surface brightness appears to
be regular and slightly elliptical at $r\gtrsim 70$\arcsec ~($50$~kpc).  To the
South, at about 60\arcsec ~($42$~kpc) from the X-ray peak, we notice the
presence of a sharp surface brightness edge similar, to those
observed in other clusters of galaxies (e.g., Markevitch et al.\ 2000,
Vikhlinin, Markevitch, \& Murray 2002a; Mazzotta et al.\ 2001).  Finally, the
X-ray image shows complex structure in the innermost $\approx
50$\arcsec ~($35$~kpc) region.  Below we study these features in detail.


 {\footnotesize
\begin{table*}[t]
\centering
\caption{Result of fitting the central 250\arcsec ~ cluster X-ray image 
 with a double 2D-elliptical-$\beta$-model}
\label{table:1}
 \noindent
 \begin{tabular}{ l c c c c c c} 
 \hline \hline
 & & & & & & \\
 Component &$x,y$ & $A$ & $r_{c}$ & $\beta$ & $\epsilon$ & $\theta$  \\
           &(J2000)   & (cnt arcsec$^{-2}$) & (arcsec) & & &(deg)\\ 
 \hline
 & & & & & & \\
 Narrow   &(03:38:40.9; +09:57:57.5){\dag} & $13.2 \pm 0.2 $& $85 \pm 1$ &$2.45\pm 0.04$ & $0.12 \pm 0.008$ & $33 \pm 6$  \\
 & & & & & & \\
 Extended &(03:38:40.4; +09:58:31.7){\dag} & $5.05\pm 0.06$ & $64\pm 1$ & $ 0.571 \pm 0.004$ & $0.13\pm 0.01$ & $34\pm 4$ \\
 \hline 
 \multicolumn{7}{l}{\dag The error of the central position is $\approx$ 2\arcsec}\\ 
 \end{tabular}
 \end{table*}
 }

\begin{figure}[t]
\centerline{\includegraphics[width=0.95\linewidth]{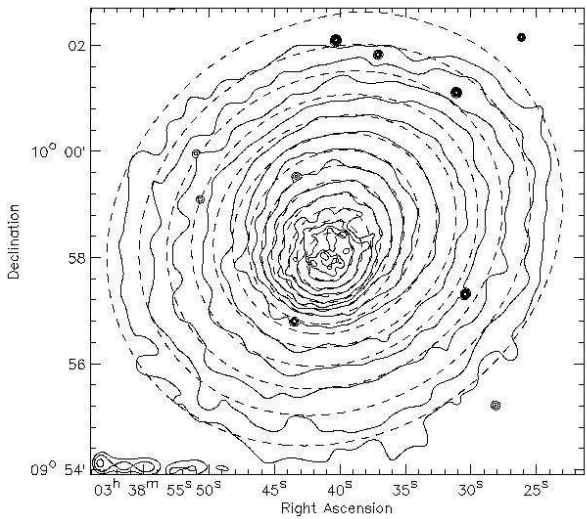}}
\caption{Comparison between the X-ray image of \2A ~ and the best fit 
double $\beta$-model. 
{\it Continuous lines}: isointensity contour levels of the cluster
image shown in Fig.~\ref{fig:x-rayimage} after adaptive smoothing.     
The image was smoothed using a Gaussian kernel with
a S/N ratio of the signal under the kernel $\ge 3$. 
The levels are spaced by a factor $\sqrt 2$. 
{\it Dashed lines}: isointensity contour levels of the best fit 
double $\beta$-model. The levels are the same as for the X-ray image.}
\label{fig:mock_image}
\end{figure}

\begin{figure}[t]
\centerline{\includegraphics[width=0.95\linewidth]{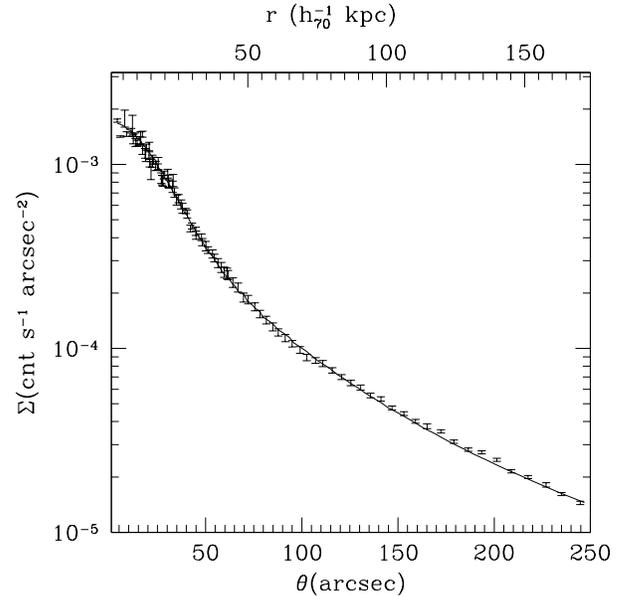}}
\caption{Comparison between the surface brightness radial profiles 
extracted from the X-ray image and the best fit double $\beta$-model.
To extract the profiles we used concentric ellipses whose 
center, ellipticity and the orientation where equal to 
the best fit values of the narrow component of the double $\beta$-model.  
Points with  errorbars and lines correspond to the profiles extracted 
from the X-ray image and the double $\beta$-model, respectively.}
\label{fig:mock_radial}
\end{figure}

 \subsection{Core structure on larger scale}\label{par:core}
 
To investigate the cluster spatial structure, and to compare it 
with previous studies,  we start with the classic approach of fitting 
the  surface brightness with a  $\beta$-model 
(Cavaliere, \& Fusco-Femiano, 1976).
Consistent with previous results, we find that this model 
does not provide an acceptable fit.
Hence, we use a double  2-dimensional  $\beta$-model defined by:
 \begin{equation}
 \Sigma=f_{narrow}+f_{extended}.
 \label{eq:beta}
 \end{equation}
 The $narrow$ and $extended$ components of eq.~\ref{eq:beta} are given by:
 \begin{equation}
 f(r,r_c,\beta,x_0,y_0,\epsilon,\theta) = A \times \Big[1+\big({r\over r_{c}}\big)^2\Big]^{-3\beta +1/2},
 \end{equation}
 where
 \begin{equation}
 r(x,y) = {\sqrt{x_{new}^2(1-\epsilon)^2 + y_{new}^2}\over(1-\epsilon)},
 \end{equation}
 and
 \begin{equation}
  x_{new} = (x - x_o)\cos(\theta) + (y - y_o)\sin(\theta).
 \end{equation}

 To perform the fit we use the SHERPA fitting 
 program\footnote{http://asc.harvard.edu/ ``Data Analysis'', ``Introduction''}.
 The cluster region with $r>250$\arcsec ~ and all 
 the strong point sources, together with the regions defined later 
 in \S\ref{par:core_small}, are masked out.
 Because the image contains a large number of pixels with few number counts
 ($< 20$~cnt) we use  the Cash (1979) statistics which, unlike the
  $\chi2$ one,  can be applied regardless of the number of counts per bin.
 The best fit values with their $90\%$ errors are reported in Table~1.

Unlike the $\chi^2$ statistics, 
the magnitude of the Cash statistics depends upon 
the number of bins included in the fit and the values of the data themselves.
This means that there is not an easy way to assess from the Cash  
magnitude how well the model fits the data.
To have a qualitative idea of the goodness of the fit, we generate 
a mock X-ray image using the 
best fit values of the double $\beta$-model and we apply 
the Poisson scatter to it.
Consistent with previous studies, we  find that globally  
the best fit double $\beta$-model describes very well  the  overall  cluster 
structure.
This is  evident in Fig.~\ref{fig:mock_image} 
where we show the isocontour levels of the model
image overlaid to the same isocontour levels of the cluster image after
adaptive smoothing. 
To better illustrate the quality of the fit  we also extract the X-ray 
surface brightness radial profiles in concentric elliptical annuli 
from both the 
cluster and the mock images as shown in  Fig.~\ref{fig:mock_radial}.   
The center, the ellipticity and the orientation of each ellipse 
are set to the best fit values of the narrow component of the 
double $\beta$-model. 
The figure clearly confirms the good quality of the fit.

The results of the fit reported in Table~1 show that  
the  two spatial components have  very different slopes,
in particular the narrow component is much steeper than the extended one.
We also notice that the profile of the narrow component derived from the 
\chandra ~ data is significantly steeper that the one obtained from the ROSAT 
PSPC (Kikuchi et al. 1999). 
The large $\beta$ value associated with the 
narrow component indicates that the cluster contains a very compact 
and dense  gas component with a quite sharp boundary 
(this is also clearly visible in  Fig.~\ref{fig:x-rayimage}). 
On the other hand, the slope of the extended component derived from the 
\chandra ~ data is significantly lower that the value derived from ROSAT  
(see e.g. Kikuchi et al. 1999, Vikhlinin, Forman, \& Jones 1999). 
Because the \chandra ~ result is obtained fitting the cluster 
surface brightness in a much smaller radial range ($r<250$\arcsec)
than the ROSAT one ($r\lesssim 1200$\arcsec),
this indicates that \2A ~ contains one or 
more radial breaks at $r>250$\arcsec, similar to the ones observed in other 
clusters like e.g. Hydra-A (David et al. 2001) and Perseus 
(Fabian et al. 2000).

Finally we note that, although the 
ellipticity and the orientation are similar, 
the centroid of the  two spatial components 
are offset by $\approx 35$\arcsec ~ with the narrow component centroid
being further south than the extended one.

Although spatial structure of the cluster is globally  well 
described by a simple double $\beta$-model, there are  two 
regions whose structure departs significantly 
from the best fit model above: i) a 60\degd ~ wide sector to the south; 
ii) the innermost $r<50$\arcsec ~ cluster region.
Below we study in details the spatial structure of these regions.

\begin{figure}
\centerline{\includegraphics[width=0.95\linewidth]{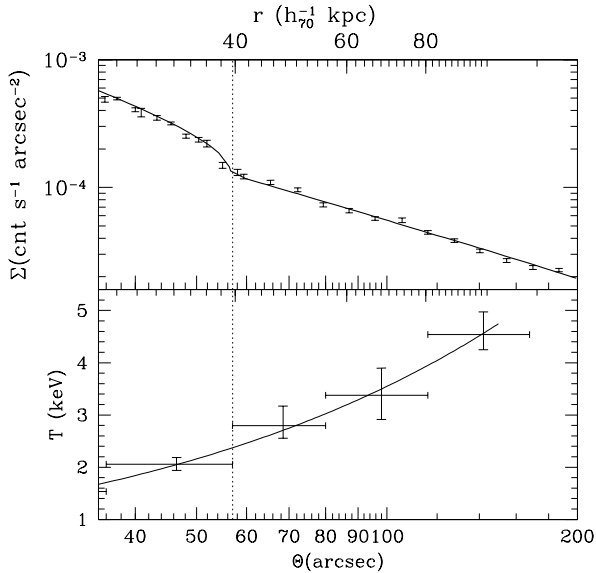}}
\caption{{\it Upper panel}) Cluster surface brightness profile
extracted from the  150\degd~ to 210\degd~ sectors.
The continuous line represents the best-fit of the discontinuous two-power law
density model. {\it Lower
panel}) Deprojected temperature profile from the 
150\degd~ to 210\degd~ sector. The continuous line is the best-fit of
 the single power law model.}
\label{fig:cold_front}
\end{figure}

\subsection{X-ray ``front''}\label{par:front}

A closer look at  Fig.~\ref{fig:x-rayimage} indicates that \2A ~ 
hosts a surface brightness drop at $r\approx 50-60$\arcsec ~ from the X-ray
peak to the south. 
This X-ray drop spans a  sector from 150 to 210~deg
and most closely resembles similar features in other clusters with dense
cores and regular morphology on larger linear scales, such as RXJ1720+26 and
A1795 (Mazzotta et al.\ 2001; Markevitch, Vikhlinin, \& Mazzotta 2001).
To better visualize this feature and to study its nature,
we extract the cluster X-ray surface brightness 
profile within the above sector, using elliptical 
annuli concentric to the brightness edge.  
The center, the ellipticity and the orientation of
each ellipses are set to the best fit values of the narrow component of the
double $\beta$-model.  
The profile is shown in the upper panel of Fig.~\ref{fig:cold_front}.  
As extensively shown in previous analysis of cold fronts in clusters
of galaxies, the particular shape of the observed surface brightness profile
indicates that the density distribution is discontinuous at some radius
$r_{jump}$. 

%
%
%


{\footnotesize
\begin{table}[t]
\centering
\caption{X-ray front density, temperature, and pressure best fit 
values\dag (errors~90\%~c.l.)}
\label{table:2}
\noindent 
\begin{tabular}{ l c c c c} 
\hline \hline
                      &        &        &   &                  \\
Quantity ($Q$)    & $s_1$  & $s_2$  & A & $r_{jump}$ (arcsec)\\
\hline
Density     ($n$) & $-1.36\pm0.04$ & $-1.20\pm0.02$ & $1.6\pm 0.02$& $56.7\pm 1$\\
Temperature ($T$) & $+0.79\pm0.03$ & $+0.67\pm0.20$ & $1.0\pm 0.20$& ...\\
Pressure    ($P$) & $-0.57\pm0.05$ & $-0.53\pm0.20$ & $1.6\pm 0.30$& ...\\
\hline
\multicolumn{5}{l}{\dag Each quantity is modeled using two power-laws
with a jump:}\\
\multicolumn{5}{l}{
$  Q= 
\left\{ 
  \begin{array}{rc} 
    (r/r_{jump})^{s_1} & r<r_{jump}  \\ 
    A_{Q}(r/r_{jump})^{s_2} & r>r_{jump}  \\ 
   \end{array}
   \right . 
$
} \\
\end{tabular}
\end{table}
}

To estimate the amplitude and the position 
of the density jump, we fit the profile
using a gas density model consisting of two power laws with index $\alpha_1$
and $\alpha_2$ and a jump by a factor $A_n$ at the radius $r_{jump}$:
\begin{equation}
  Q= 
\left\{ 
  \begin{array}{rc} 
    (r/r_{jump})^{\alpha_1} & r<r_{jump} \, , \\ 
    A_n(r/r_{jump})^{\alpha_2} & r>r_{jump} \, . \\ 
   \end{array}
   \right . 
\label{eq:5}
\end{equation}
The
fit is restricted to  radii $r=30-200$\arcsec ~ to exclude the central region
with irregular structure. We assume spherical symmetry. 
The best fit values together with the 90\% confidence level errors
are shown in Table~2. 
The best fit model (which  provides a good fit to
the data)  is shown with a dashed line in
the upper panel of Fig.~\ref{fig:cold_front}.


{\footnotesize
\begin{table}
\centering
\caption{Spatial and spectral properties of the X-ray blobs}
\label{table:3}
\begin{tabular}{ c c c c c c c} 
\hline \hline
& & & & & & \\
Region &  $r_1$ & $r_2$ &  net cnts &$n_e$ & $L_{[0.6-9]}$  & $t_c$  \\
       & (arcsec) & (arcsec) &([0.6-9]~keV) &($10^{-2}$cm$^{-3}$)  &$10^{41}\,$
erg\, s$^{-1}$
&($108\, $yr)\\ 
\hline
1 & 9.0& 4.8 & 796  &   8.9 &  5.8 & 2.0  \\
2 & 7.5& 5.6 & 1092 &  10.0 &  8.2 & 1.9    \\
3 & 5.8& 4.6 & 724  &  12.7 &  6.4 & 2.1  \\
4 & 6.6& 3.7 & 582  &   9.4 &  4.6 & 2.4  \\
5 & 7.4& 5.1 & 405  &   6.4 &  2.7 & 2.4  \\
6 & 8.7& 5.1 & 684  &   7.6 &  7.5 & 4.1  \\
7 & 4.4& 4.3 & 223  &   8.3 &  1.8 & 3.3  \\
8 & 5.2& 5.2 & 312  &   7.2 &  2.4 & 2.8  \\
\hline 
\end{tabular}
\end{table}
}

%
%
%

\begin{figure}
\centerline{\includegraphics[width=0.95\linewidth]{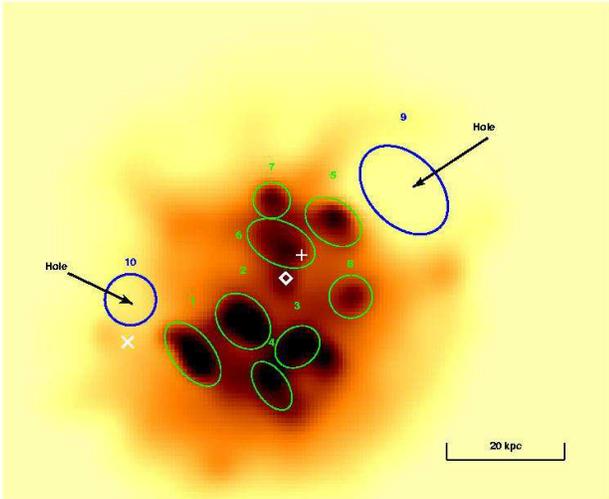}}
\caption{Zoom in of Fig.~\ref{fig:x-rayimage} after adaptive smoothing made
using a Gaussian kernel with a S/N ratio of the signal under the kernel $\ge 3$.
The green ellipses labeled from 1 to 8 identify 8 X-ray bright extended regions or blobs.
The two blue ellipses, labeled 9 and 10, identify the Western and the Eastern X-ray holes or 
cavities. The white square and cross, indicate the position of the cluster D galaxy
and its companion, respectively. The white X symbol indicate the position of the other cluster central galaxy see (Fig.~\ref{fig:hst_image}).}
\label{fig:im_zoomed}
\end{figure}

%
{\footnotesize
\begin{table*}
\centering 
\caption{Comparison of Spectral Fits for the innermost 250\arcsec \ region}
\label{table:4}
\begin{small}
\begin{tabular}{ l c c c c c c }
\hline \hline
 & & & & & & \\
Model  & $N_H$  $(10^{21}$\ cm$^{-2})$ & $T_{1}$ (keV) & $T_{2}$ (keV) & $Z/Z_{\odot}$ & $\dot {M}$ $(M_{\odot}{\rm yr}^{-2})$ & $\chi^2/d.o.f.$ \\
\hline
 & & & & & & \\
TBABS$\times$MEKAL& $1.6$ & $2.7$ & ... & $0.59$ & ... & 843/352 \\
& $(1.6^{+0.02}_{-0.02})$ & $(2.6^{+0.05}_{-0.05}) $ & ... & $(0.54^{+0.04}_{-0.04})$ & ... & (484/297) \\
 & & & & & & \\
TBABS$\times$[MEKAL+MEKAL]& $1.71_{-0.05}^{+0.06}$ & $3.3_{-0.2}^{+0.9}$ & $1.4_{-0.1}^{+0.3}$ & $0.55_{-0.05}^{+0.05}$ & ... & 643/350 \\
      & $(1.71_{-0.07}^{+0.08})$ & $(3.0_{-0.2}^{+0.9})$ & $(1.4_{-0.2}^{+0.3})$ & $(0.53_{-0.06}^{+0.06})$ & ... & (393/295) \\
 & & & & & & \\
TBABS$\times$[MEKAL+MKCFLOW]&$2.0^{+0.1}_{-0.1}$&$2.90^{+0.1}_{-0.1}$&...& $0.68^{+0.04}_{-0.04}$&$247^{+37}_{-34}$&658/351 \\ 
           &$(2.0^{+0.1}_{-0.1})$&$(2.78^{+0.06}_{-0.06})$&...& $(0.60^{+0.04}_{-0.04})$&$(192^{+37}_{-34})$&(387/296)\\ 
 & & & & & & \\
TBABS$\times$[MEKAL &$0.9_{-0.2}^{+0.2}${\dag}& $3.0_{-0.1}^{+0.1}$&...&$0.68_{-0.02}^{+0.02}$ & $266_{-36}^{+35}$ & 658/351 \\
+ZTBABS(MKCFLOW)] &$(1.3_{-0.4}^{+0.4})${\dag}& $(2.8_{-0.1}^{+0.1})$&...&$(0.61_{-0.04}^{+0.05})$ & $(249_{-60}^{+65})$ & (393/296) \\
\hline 
\multicolumn{7}{l}{\dag Intrinsic absorption}\\
\multicolumn{7}{l}{Values in parentheses are from fits with the excluded
  1.4--2.2 keV energy interval.}\\
\end{tabular}
\end{small}
\end{table*}
}

%
%
\subsection{Core structure on smaller scales}\label{par:core_small}

To study the cluster morphology on  smaller scales  
we adaptively smooth the cluster image as shown in Fig.~\ref{fig:im_zoomed}. 
The image was smoothed using the 
CIAO\footnote{http://asc.harvard.edu/ ``Data Analysis''} 
tool CSMOOTH with a Gaussian kernel and
a S/N ratio of the signal under the kernel $\ge 3$. 

The image clearly shows a very complex structure 
with a number of blobs of X-ray excess emission. 
The unsmoothed image shows that these blobs are not necessarily
distinct  but may form a filamentary structure.  
Nevertheless, for the purpose of our analysis we assume
that the X-ray blobs are ellipsoid.
We select 8 regions as shown in  Fig.~\ref{fig:im_zoomed}.
Each ellipse indicates the region where the surface brightness  
drops by almost a factor two with respect to the peak of its respective 
X-ray blob.
To measure the volume occupied by each blob we make the simple assumption
that the radius of the ellipsoid along
the line of sight is equal to the minor radius of the ellipse $r_2$. 
In  Table~3 we report the major and the minor radii corresponding to each
region shown in  Fig.~\ref{fig:im_zoomed}.
The dimensions of the X-ray blobs span from $4$\arcsec ~ to $9$\arcsec ~ (from 2.8 to
6.3~kpc).  
For each blob we calculate the mean  electron density using the 
following simple procedure:  using the method described 
later in \S~\ref{par:spectral},
we extract the spectra from both  the blob region and  
from a cluster region just outside it. 
Using the latter spectrum as background, we fit the blob
spectrum with a thermal model, hence  deriving the emission measure.   
Finally, the gas density is calculated from the  emission measure
under the simple assumption that it is constant within the blob.
The net number of counts and the mean gas densities obtained using the above
procedure are reported Table~3. 
For completeness in the same table we add the unabsorbed luminosity in the 
(0.6-9)~keV band in the source rest
frame and the cooling time (see \S\ref{par:cavities} below).

Beside  the X-ray blobs, we also note a prominent X-ray hole 
about $30$\arcsec ~ ($21$~kpc) North-West from the X-ray peak, and a less
prominent one at $\approx 40$\arcsec ~ ($38$~kpc) to the East.
To check the statistical significance of the Eastern hole, we extract
the count rates from a circular region  in the hole ($r=6.5^\second$;
see Fig.~\ref{fig:im_zoomed}) and the count rates in two similar regions to
the East and to the West of the depression along the isocontour level
 defined by the best-fit double
$\beta$-model.  The absorbed flux from the hole is $\approx 20\%$ lower
than the flux from the neighboring regions at a $4.5\sigma$ confidence
level. As shown later in \S\ref{par:other}, the cluster image 
may also  present  other X-ray holes at larger radii.
The previous identification of the Eastern hole from ROSAT
HRI observations with absorption by 
a cluster member galaxy by Sarazin et al.
(1992, 1995) is not supported with these higher resolution 
images (see Fig.~\ref{fig:im_zoomed}).

\begin{figure*}
\centerline{\includegraphics[width=0.95\linewidth]{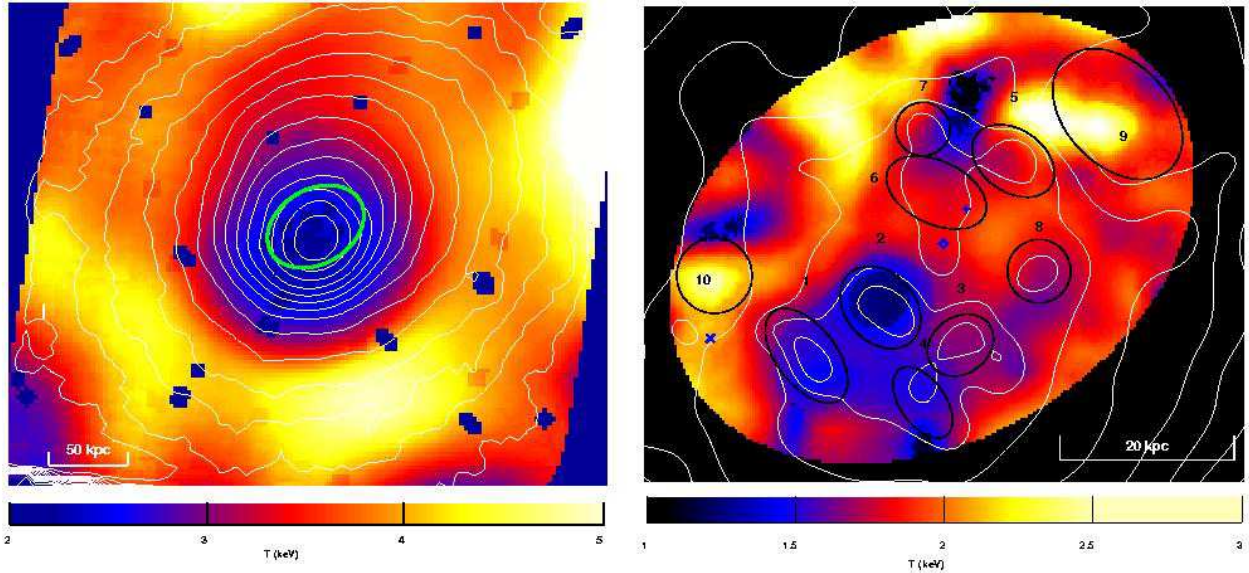}}
\caption{Temperature map of \2A ~ overlaid on the 0.3-9~keV ACIS 
brightness contours.
{\it Left panel}) 
This temperature map was obtained by
smoothing the images using a Gaussian filter with $\sigma=15^\second$.
The green ellipse corresponds to temperature map on
the right panel. 
Blue holes indicate the position of the point sources that were 
masked out. The isocontours are obtained from the image after a
Gaussian smoothing with $\sigma= 15^\second$. 
Levels are spaced by a factor $\sqrt 2$. 
{\it Right panel}) Zoom in, high resolution temperature map 
from the cluster  inside the green ellipse on the left panel. 
This temperature map was obtained by
smoothing the images using a Gaussian filter with $\sigma=4^\second$.
Isolevels are the same as in the continuous lines in Fig~\ref{fig:mock_image}.
The black ellipses labeled from 1 to 10 identify  the regions 
defined in Fig~\ref{fig:im_zoomed}. The blue square, cross, and X symbols
indicate the position of the galaxies as in Fig.~\ref{fig:im_zoomed}.}
\label{fig:temp_map}
\end{figure*}

%
%
%
\section{Spectral Analysis}\label{par:spectral}

We use the CIAO software to extract spectra. The spectra are grouped to
have a minimum of 20 net counts per bin and fitted using the XSPEC package
(Arnaud 1996).  The position-dependent RMFs and ARFs are computed and
weighted by the X-ray brightness over the corresponding image region using
``calcrmf'' and ``calcarf'' tools\footnote{A. Vikhlinin 2000,
(http://asc.harvard.edu/ ``Software Exchange'', ``Contributed Software'').}.
To account for the absorption caused by molecular contamination of the ACIS
optical blocking filters\footnote{ http://asc.harvard.edu/ ``Instrument \&
Calibration'', ``ACIS'', ``ACIS QE Degradation''}, all the spectral model we
use are multiplied by the \textsc{acisabs} absorption model\footnote{Chartas
\& Getman 2002, http://www.astro.psu.edu/users/chartas/xcontdir/xcont.html}
that models the above absorption as a function of time since launch.  The
parameter \textsc{Tdays} is set to 410, while the others are left to the
default values, namely \textsc{norm}$=7.22\ 10^{-3}$,
\textsc{tauinf}$=0.582$, \textsc{tefold}$=620$, \textsc{nC}$=10$,
\textsc{nH}$=20$, \textsc{nO}$=2$, \textsc{nN}$=1$.

Spectra are extracted in the 0.6-9.0~keV band in PI channels.  At present,
the ACIS response is poorly calibrated around the mirror Ir edge, which
results in frequently observed significant residuals in the 1.4--2.2 keV
energy interval and high $\chi^2$ values.  We tried excluding this energy
interval and found that the best-fit parameter values and confidence
intervals do not change considerably while the $\chi^2$ values become
acceptable, as will be seen below.  We therefore elected to use the full
spectral range.

\subsection{Average Temperature}\label{par:average}

To determine the average spectral properties of this observation, we extract
the overall spectrum from the innermost circular region of $r=250\arcsec$
($175$~kpc) centered on the X-ray peak and fit it with different models as
listed in Table~4.  In the following the absorption of X-rays due to the
interstellar medium has been parametrized using the T\"ubingen-Boulder model
(\textsc{tbabs} in XSPEC v. 11.1; Wilms, Allen \& Mc-Cray 2000).
Furthermore the metallicity refers to the solar photospheric values in
Anders \& Grevesse (1989).  As a first attempt, we use an absorbed single
temperature thermal model (\textsc{tbabs}$\times$\textsc{mekal}; 
see e.g. Kaastra, 1992;
Liedahl, Osterheld, \& Goldstein, 1995; and references therein).  We
find that this model does not reproduce the spectrum.  We also use an
absorbed two-temperature model
(\textsc{tbabs}$\times$[\textsc{mekal}+\textsc{mekal}]) with the
metallicity of the two components linked together.  The hydrogen equivalent
column density and the temperature, the metallicity, and the normalization
of each thermal component are left free to vary.  
This model provides a much better fit. In particular, if we exclude  
the energy band near the Ir edge  from the spectral analysis, 
we find that the model is statistically acceptable
(see Table~4).
We find that the equivalent
column density is consistent with the galactic value
($N_H=1.78\times10^{21}$~cm$^{-2}$; Dickey \& Lockman 1990). The
temperatures of both the hot and the cool components, as well as the
metallicity, are consistent with the values obtained from the combined
analysis of ASCA GIS and SIS spectra extracted from the innermost
$r=120\arcsec$ region (Kikuchy et al. 1999).  We finally fit the spectrum
using two models with cooling flows.  The first is an absorbed thermal
model plus a cooling flow component
(\textsc{tbabs}$\times$\textsc{[mekal+mkcflow]}).  In the latter we
assumed that the cooling flow component itself is intrinsically absorbed by
uniformly distributed amount of hydrogen at the cluster redshift
(\textsc{tbabs}$\times$\textsc{[mekal+ztbabs(mkcflow)]}, in XSPEC v. 11.1).
The higher temperature and metallicity of the cooling flow component are
fixed to be equal to the ones of the thermal component.  In the first
cooling flow we leave the absorption free to vary. In the intrinsically
absorbed cooling flow model we fix the overall absorption to the Galactic
value and let the intrinsic $N_H$ free to vary.  We find that both
models give temperatures, metallicities and the mass deposition rates
consistent within the errors.  It is worth noting that the cooling flow
models give a minimum $\chi^2$ value similar to the two-temperature model
which, as said before, is quite high compared to the degree of freedom
(d.o.f).

\subsection{Temperature Map}\label{par:tmap}

To determine the temperature map we used the same technique used for A3667
(Vikhlinin, et al. 2001a; Mazzotta, Fusco-Femiano, Vikhlinin 2002).
We extracted images in 10 energy bands,
$0.80-1.07-1.43-1.68-1.82-2.16-2.35-2.88-3.75-5.83-9.00$~keV, 
subtracted background,
masked out point sources, divided by the vignetting maps
and  smoothed the final images.  
We determined the
temperature by fitting the 10 energy points to a single-temperature
absorbed \textsc{mekal} plasma model with metal
element abundance fixed at  $Z=0.6\ Z_\odot$.  We left as free parameters 
the equivalent absorption column density, the temperature, 
and the normalization.

To study the cluster temperature structure at larger radii 
(where the cluster is less bright), we derive first a relatively low 
spatial resolution temperature map. This is obtained by smoothing the 10 
images above using a Gaussian filter with $\sigma=15^\second$.
The ``low-resolution'' temperate map, overlaid on the 0.3-9~kev ACIS brightness
contours, is shown in the left panel of Fig.~\ref{fig:temp_map}.  
Like the surface brightness,  the projected temperature map is not
azimuthally symmetric.  The gas is substantially cooler in the cluster
center and shows a strong positive temperature gradient which varies
with position angle. The temperature gradient seems to be higher and
lower to the Southern and Northern sectors, respectively.  It is interesting
to note that the gradient seems to be maximum somewhere in the south sector,
where we observe the surface brightness edge.

To study the thermal properties of the  X-ray blobs and 
cavities discussed in \S\ref{par:core_small}, we produce a second 
temperature map of the innermost central region
using a much higher spatial resolution. 
This region is identified by the green ellipse in 
left panel of Fig.~\ref{fig:temp_map}.
The ``high-resolution'' temperature map, overlaid on the 
0.3-9~kev ACIS brightness contours, is shown in the right panel 
of Fig.~\ref{fig:temp_map}.  
The map is obtained by smoothing the 10 images above
using a very narrow Gaussian filter with $\sigma=4^\second$.
For clarity we plot ellipses numbered from 1 to 10
corresponding to the 8 X-ray blobs and 2 cavities defined 
in the \S\ref{par:core_small}, above (see also Fig.~\ref{fig:im_zoomed}).
The high-resolution temperature map shows a very complex structure.
In particular the cavities and the X-ray blobs seem to have 
quite  different projected temperatures. 
We discuss this aspect in \S\ref{par:cavities} below.

\begin{figure}
\centerline{\includegraphics[width=0.95\linewidth]{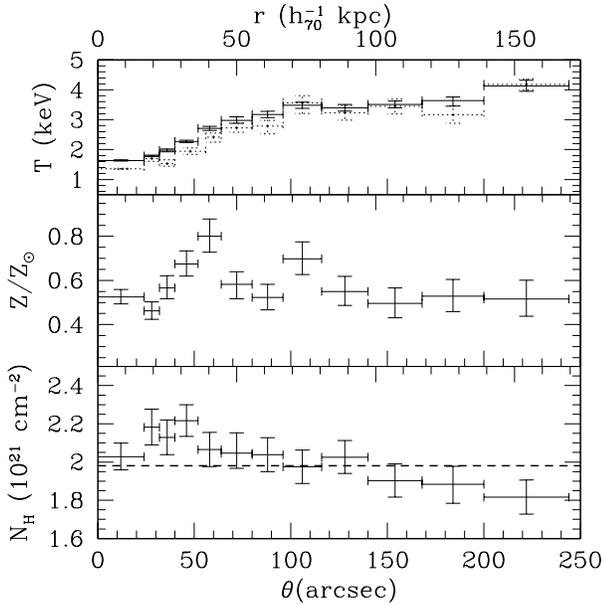}}
\caption{Emission weighted temperature, metallicity, and equivalent column density 
radial profiles of \2A ~.
The spectra are extracted in concentric elliptical annuli. 
Center, ellipticity and orientation of each ellipses 
is fixed to the best fit values of the narrow component of the 
double $\beta$-model as shown in  Fig.~\ref{fig:mock_radial}. The
dotted crosses in the upper panel 
indicate the results obtained by deprojecting the
temperature profiles in the assumption of spherical symmetry.
The dotted line in the bottom panel 
is $N_{H}$ value obtained by fitting a constant to the $N_H$ profile. 
Error bars are $1\sigma$ confidence level.}
\label{fig:profile_0_360}
\end{figure}

\begin{figure}
\centerline{\includegraphics[width=0.95\linewidth]{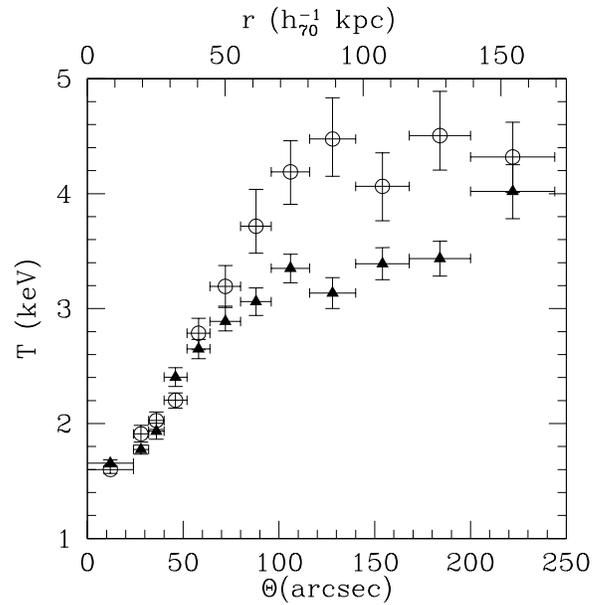}}
\caption{Comparison of the temperature profiles extracted from
the   Northern half (filled triangles) and Southern half (open
circles) of the cluster.}
\label{fig:profile_0_180}
\end{figure}

\subsection{Radial temperature gradients}\label{par:tprofile}

We first measure the overall spectral radial properties  
extracting spectra in concentric elliptical annuli. 
Center, ellipticity and orientation of each ellipse 
are fixed to the best fit values of the narrow component of the 
double $\beta$-model as shown in  Fig.~\ref{fig:mock_radial}. 
The radii are choose in order to have $\approx 20,000$ net counts
per spatial bin.
Each spectrum is fitted with an absorbed one-temperature  
\textsc{mekal} model.  
Equivalent hydrogen column density, temperature, metallicity,
and normalization are left free to vary.
We get an acceptable fit for all the annuli but the innermost 
one.
The resulting temperature, metallicity, and $N_H$  profiles,
with error bars at $1\sigma$ level, are shown in the upper, middle, and lower 
panel of
Fig.~\ref{fig:profile_0_360},  respectively 
(errors bars are $1\sigma$ level).
The temperature profile shows a positive 
gradient within the innermost  $50-60 h_{70}^{-1}$~kpc
with the temperature that goes from 
 $\approx 1.5$~keV to $\approx 3.5$~kev.
At larger radii the temperature profile shows a much flatter profile
consistent with being constant. Only the last bin at 
$150 h_{70}^{-1}$~kpc indicates a possible temperature 
increment at larger radii.  
Although the cluster is clearly not spherical symmetric we try
to deproject the overall temperature profiles assuming it actually is.
The deprojected  profile is reported as dotted crosses in 
the upper panel of Fig.~\ref{fig:profile_0_360}. 
We notice that the deprojected 
 temperature of the innermost bin is $T=1.36\pm 0.02$~keV.

The projected metallicity is  
consistent with the constant value $Z\approx 0.5 Z_{\odot}$ 
at radii $> 50-60 h_{70}^{-1}$~kpc. 
At lower radii it shows strong positive gradient 
with a maximum of $\approx 0.8 Z_{\odot}$ at $40 h_{70}^{-1}$~kpc.
 Similar metallicity gradients have been observed 
in other clusters, e.g. M87 (Molendi \& 
Gastaldello, 2001),  Persus (Schmidt, Fabian and Sanders, 2002), 
Abell~2052 (Blanton, Sarazin, McNamara, 2002), A2199 (Johnstone et al. 2002)
and Centaurus (Sanders \& Fabian 2002). 
For some of them, however, 
it has been shown that the  gradient is an 
artifact  induced by the inadequate 
modeling of the spectra with a single temperature component where  
a multi-temperature one is actually required. This effect is 
known as metallicity bias  (see Buote 2000). 
It is interesting to note, however, that if we  deprojected 
the metallicity profile assuming spherical symmetry
we find that the profile is consistent with the projected one.
This  either indicates that the spherical symmetry assumption 
 is inadequate  or that the gradient is real.
Recently Morris \& Fabian (2003) suggested that apparent
 off-center peaked  metallicity profiles
may form as result of the  thermal evolution  
of intracluster medium  with a non-uniform metal distribution.

Finally, as shown in Fig.~\ref{fig:profile_0_360}, panel c), 
we find that the 
equivalent column density profile is consistent with a constant value of
$1.98\times 10^{21}$~cm$^{-2}$ which seems to be 10\% higher 
than the galactic value.

As discussed in \S\ref{par:tmap}, 
the clusters temperature distribution is strongly asymmetric.
To verify the statistical significance of this asymmetry, 
we extract the spectra using the photons from the
Northern or the Southern sectors. 
Spectra are extracted  in the  elliptical annuli, with the same binning 
as before. 
We find that, although both the abundance and metallicity profiles
from the North and the South sectors are consistent within the errors,  
 the temperatures profiles are significantly different.
 Fig.~\ref{fig:profile_0_180} clearly shows that 
the temperature gradient in the North sector is shallower
than the gradient in the South sector.
Moreover, the  region of constant temperature in the South sector
is a factor 1.3 hotter than that in the North sector.  

\subsection{Temperature profile of the ``front''}\label{par:tprofile_NS}

In  \S\ref{par:front} we show that the cluster has a surface brightness edge
 similar to the cold fronts observed in other clusters of galaxies.
One peculiar characteristic of these cold fronts is that they have 
a temperature discontinuity (jump) over the front itself.
In this paragraph we study  
the thermal properties of the X-ray ``front'' in \2A .
To do this we estimate the 3D temperature profile of the front by 
 deprojecting the temperature profile from the cluster sector whose angles 
are from 150\degd~ to 210\degd. The 3D temperature profile is shown in 
the lower panel of Fig.~\ref{fig:cold_front}.  
We notice that the temperature profile has  a quite steep gradient.
Moreover, unlike the cold fronts observed in other clusters of galaxies, 
it does not show a clear temperature discontinuity at the front.
At  first we fit the  temperature profile in the range from
30\arcsec ~ to 150\arcsec ~ with a  single power-law model 
$T\propto \theta^{\delta}$. The model provides a good fit with
$\delta=0.72\pm0.02$, as shown  in  Fig.~\ref{fig:cold_front}.
We notice that the data may also be consistent with a small 
temperature jump at the front. 
To constrain  the amplitude of this possible jump, we
fit the data using a two-power law model with 
a jump $A_{T}$ at the position of the front (see eq~\ref{eq:5}) and we 
report the result in  Table~2.
From the table we see that  $A_{T}=1\pm0.2$.
Using the pressure and the density models above, we calculate
the pressure profile near the front region. The best fit model
together  with the 90\% errors are reported
in Table~2. 
 The observed  density and  temperature   
discontinuity at the front  result in   
corresponding  pressure discontinuity of a factor 
$A_{P}=1.6\pm 0.3$.

\begin{figure}
\centerline{\includegraphics[width=0.95\linewidth]{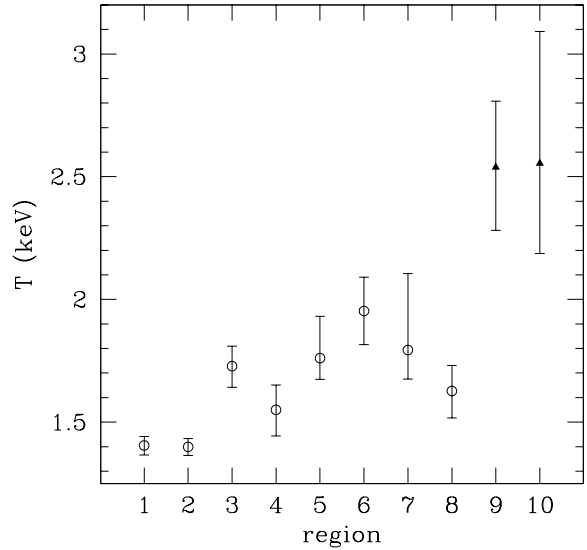}}
\caption{Projected temperature from  the 8 
regions defined in Fig.~\ref{fig:im_zoomed}. The
two filled triangles give the projected temperature from the Western and 
Eastern
 holes, respectively. 
Error bars are $90\%$ confidence level.}
\label{fig:hardness}
\end{figure}

\begin{figure}
\centerline{\includegraphics[width=0.95\linewidth]{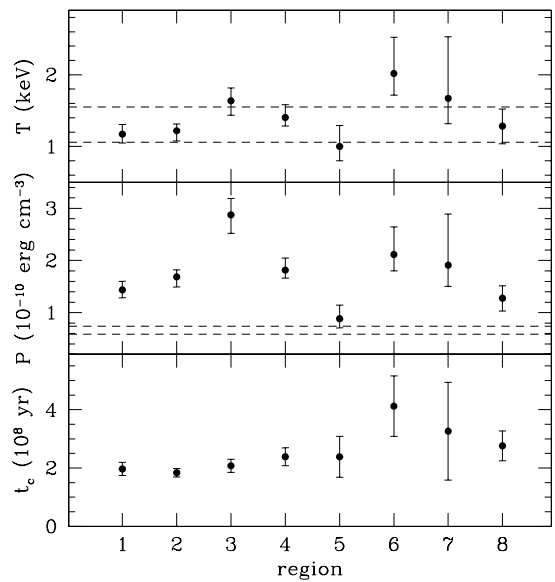}}
\caption{{\it Upper panel}) Deprojected temperature associated with each of 
the X-ray 
blobs region defined in  Fig.~\ref{fig:im_zoomed}. The dashed lines are
the 90\% error measurement on the deprojected  temperature of the ambient gas.
{\it Middle panel}) Thermal pressure associated with the same blobs as in the Upper panel.  The dashed lines are
the 90\% error measurement of the thermal pressure of the ambient gas.
 {\it Lower panel}) Cooling time associated with the same blobs as in the 
Upper panel.}
\label{fig:pressure}
\end{figure}

\subsection{Spectral properties of the X-ray blobs and cavities}\label{par:cavities}
 
As discussed in \S\ref{par:tmap}, the innermost region of \2A ~ 
has a very complex temperature 
structure. 
In particular, the temperature map shows 
that  the cavities and the X-ray blobs have 
quite different projected temperatures (see Fig.~\ref{fig:temp_map}). 
To better study the thermal properties of these structures,
we  extract spectra for all the elliptical regions shown in the Figure
and fit them with an absorbed thermal model.    
We verify if there is any  excess absorption by
fixing the metallicity to Z=0.6 solar and letting both $N_H$ and 
the temperature  to vary freely.  
As we find that the resulting $N_H$ values are all consistent with the cluster
mean value, we fix $N_H$ to this value and re-fit the model to measure the
temperature.  The projected temperatures
together with the 90\% error bars, are shown in Fig.~\ref{fig:hardness}.
Consistently with the right panel of 
Fig.~\ref{fig:temp_map}, we find that the central X-ray structures have
different projected temperatures.  
In particular we notice that region 1 and 2 are significantly 
cooler than region 6 and 7.
Furthermore, we notice that the holes are significantly hotter than any 
other X-ray blobs.
It is worth noting that the observed higher temperature from the holes is 
likely to be the result of a  projection effect.

Projection could
also be responsible for the observed difference in the projected temperature
of the X-ray blobs.  To verify this hypothesis one should find a way to
disentangle the contribution to the spectrum due to the blob itself from
that due to the gas along the line of sight.  To perform this task we assume
that the contribution to the spectra due to the gas along the line of sight
is almost the same within the innermost 25\arcsec ~ region (where all the
blobs are).  This is justified by the fact that the surface brightness
profile within this region does not vary more than a factor $\approx
1.5$. We then extract a cluster spectrum using a region just outside the
blobs that will be used as background for the spectrum of each blob.  
The first important result of this exercise is that the final spectra clearly
shows the presence of Fe-L emission line complex. 
This proves, without doubt, that the
emission from these blobs is indeed mostly, if not all, thermal. 
We fit these spectra using a
single-temperature \textsc{mekal} model leaving as a free parameter only the
temperature and the normalization.  The result is shown in the upper panel
of Fig.~\ref{fig:pressure}.  In the same figure we add, as dashed lines, the
90\% error temperture estimate of the ambient gas. This estimate cames from
the fit of the blob-free cluster spectrum, used before as background, with
a two-temperature \textsc{mekal} model with the metallicity of the two thermal
components linked together.  To compensate our ignorance of the actual 3D
cluster structure, we leave the normalization of both thermal components
free to vary.  We assume that the temperature of the ambient gas coincides
with the one of lower temperature component. We believe that this represents
a good assumption as the value we find is consistent with the temperature of
the innermost bin obtained using a spherical symmetric deprojection model
(see upper panel of Fig.~\ref{fig:profile_0_360}).

Unlike the emission weighted temperatures, 
the  deprojected ones have a much lower scatter. Furthermore, all regions,
but the  6th, have temperatures consistent with the 3D temperature 
of the ambient gas. 

Using the blob density measurements reported in Table~3, it is now possible
to estimate the thermal pressure associated with each blob.  This is shown
in the central panel of Fig.~\ref{fig:pressure}. In the same figure we add,
as dashed lines, the 90\% error ambient pressure. The blobs, except
region 5, have  thermal pressure values  significantly higher than
the ambient thermal pressure. This unexpected finding will be
discussed below.

Using the temperature and the density of each blob we also estimate
their cooling times $\tau_c \equiv 5 P/ 2\epsilon$, where $P$ and 
$\epsilon$ are the blob's pressure and the emissivity, respectively. 
These, together with the 90\% error bar,
 are  shown in the lower panel of
Fig.~\ref{fig:pressure}. We notice that all blobs
have similar cooling times which are consistent within the errors
(but see region 6).
For practical reasons we report also the blob cooling times in column 7 
of Table~3.

For completeness  we  estimate  the X-ray blob unabsorbed luminosities
in the (0.6-9.0)~keV in the source rest frame, as shown in  
in column 6  of Table~3. 
\begin{figure}
\centerline{\includegraphics[width=0.95\linewidth]{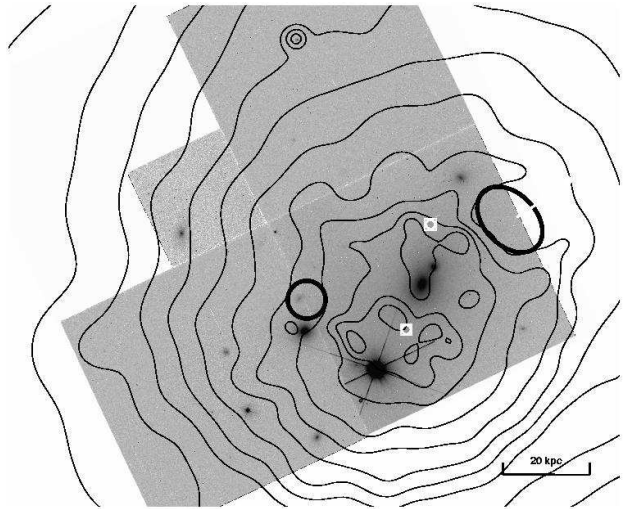}}
\caption{HST image of the central region of \2A ~ superposed on contours of 
the X-ray image after adaptive smoothing. The circles in square symbols 
indicate the best fit central position of each of the components of the 
double $\beta$-model. The Northern and the Southern
symbols indicate the position of the Extended and Narrow component, 
respectively. The thick ellipses indicate the position of the Western 
and Eastern holes, respectively. The X-ray excess regions labeled 1 to 8 in 
Fig.~\ref{fig:im_zoomed} can be easily identified by the small  
quasi-elliptical isocontour regions. Notice that none of these regions have
an optical counterpart.}
\label{fig:hst_image}
\end{figure}

\begin{figure*}
\centerline{\includegraphics[width=0.95\linewidth]{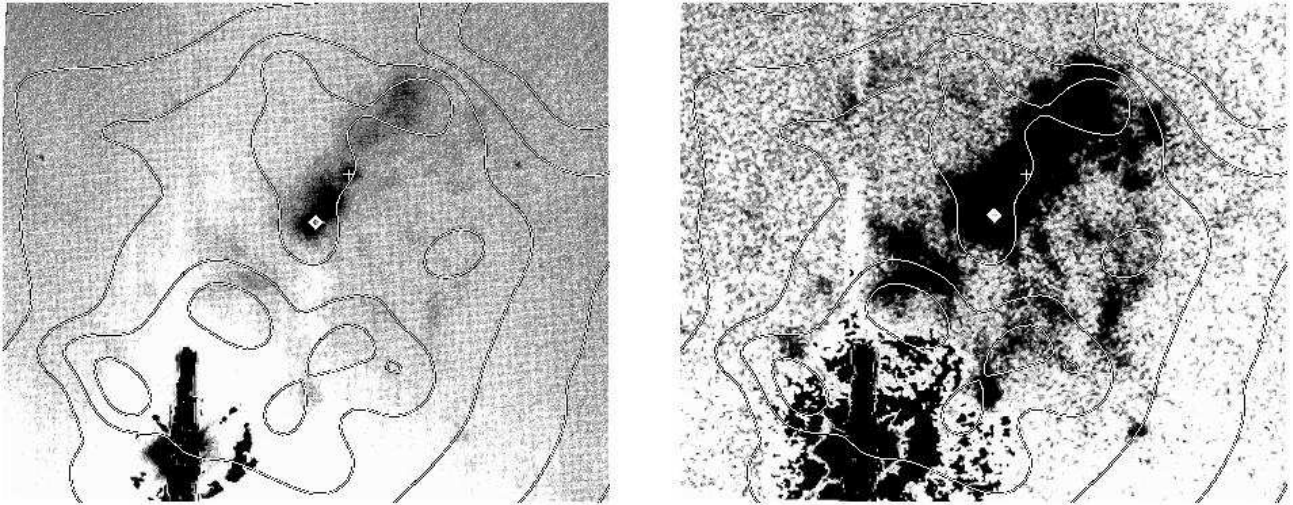}}
\caption{H${\alpha}+$[N\textsc{ii}] image of the central 
$60^{\prime\prime}\times 70^{\prime\prime}$ region of \2A ~ (from  
Romanishin \& Hintzen 1988, Fig.~3) with   X-ray  
contours overlaid. The X-ray contours are the same as 
in Fig.~\ref{fig:hst_image}. 
To better visualize the faintest line features 
we show the H${\alpha}+$[N\textsc{ii}] image in both high and low contrast 
(left and right panels, respectively).
The square and the cross symbols indicate the position of the central D 
galaxy and its companion, respectively.
The X-ray excess region labeled 1 to 8 in 
Fig.~\ref{fig:im_zoomed} can be easily identified by the small  
quasi-elliptical isocontour regions.
The Western X-ray cavity is located in the upper right hand corner.
Note that  the large, bright artifact in the lower left of the 
image is a saturated star.}
\label{fig:halpha}
\end{figure*}

\begin{figure}
\centerline{\includegraphics[width=0.95\linewidth]{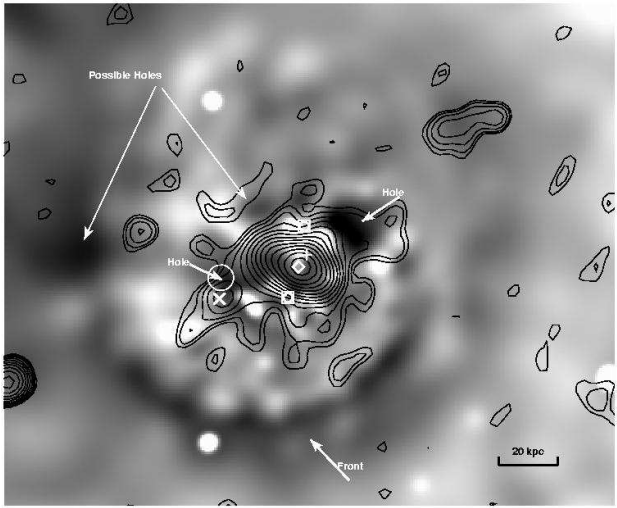}}
\caption{
Percentile variation of the cluster X-ray image (after adaptive smoothing) 
with respect to the best fit double $\beta$-model 
superposed on contours of the 1.515 GHz 
radio image (from Sarazin, Baum, \& O'Dea, 1995). 
The darker and lighter regions indicate  negative and positive variation
$\ge |20|\%$, respectively. The arrows indicate the position of the Western and Eastern holes and the position of the Front.
The square and the cross symbols indicate the position of the central D 
galaxy and its companion while the X symbol indicates 
the position of the third cluster galaxy in the HST field 
(see Fig.~\ref{fig:hst_image}). 
Finally the  circles in squares symbols indicate 
the best fit central position of each of the components of the 
double $\beta$-model. }
\label{fig:percentile}
\end{figure}

\section{Properties at other wavelengths}\label{par:other}

We conclude the  analysis of \2A ~ by comparing the cluster  properties in
the X-rays to those at other wavelengths.

In Fig.~\ref{fig:hst_image} we show an HST image of the central 
region of \2A ~ superposed on contours of 
the X-ray image after adaptive smoothing. 
The optical image shows one central D galaxy
 (PGC 013424, Paturel et al. 1989) 
 with a companion 
smaller galaxy 7\arcsec ~to the North-West and another elliptical galaxy at
41\arcsec ~to the East. 
In Fig.~\ref{fig:hst_image} 
the X-ray blobs, defined previously in \S\ref{par:core_small}
(labeled 1 to 8 in Fig.~\ref{fig:im_zoomed}), 
can be easily identified matching the small  
quasi-elliptical isocontour regions.
Apart from 
one X-ray blob  coincident with the 
cluster D galaxy  none of the other blobs  
have optical counterparts. This strongly suggests that the 
X-ray blobs are produced by some kind of hydrodynamic process. 

For completeness  in Fig.~\ref{fig:hst_image} we report the centroid 
position of the narrow and extended $\beta$-model components. 
It is worth noting that the D galaxy position does not coincide
with either of these centroids but rather lies in between.

The central galaxy of 2A~0335$+$096  has been found to
have a number of peculiar properties over the past 2 decades. 
Initially it was classified as a Seyfert 1.5 or Liner by Huchra, 
Wyatt \& Davis (1982) but more detailed optical observations
by Romanishin \& Hintzen (1988) found a very extended optical emission
line region (H$\alpha+$[N\textsc{ii}])  
to the North-East of the galaxy (extended over 30~kpc) 
and the central regions of the galaxy are anomalously blue
indicating recent star-formation (see Fig.~\ref{fig:halpha}).
Recently a number of authors have shown that  some of
the extended  H$\alpha$ features observed  
in the central region of a number of clusters correlate with similar
structures of X-ray excess  
(e.g. A1795, Fabian et al. 2001; RX J0820.9+0752, Bayer-Kim et al. 2002; 
Virgo, Young, Wilson, \& Mundell 2002).

To study the correlation between the X-ray blobs and the line emission in
Fig.~\ref{fig:halpha} we show the narrow band H$\alpha+$[N\textsc{ii}] image
from  Romanishin \& Hintzen (1988)  with the X-ray isocontours superimposed.
The accuracy of the alignment is better than $3^{\prime\prime}$.
There is no direct correlation between the X-ray and H$\alpha+$[N\textsc{ii}] structures
but there is a tendency for the two to lie close to each other. However,
the statistical significance of this tendency is difficult to judge.
Note the X-ray blobs labeled 1 and 4 lie  
too close to the saturated star to the South East of the H$\alpha+$[N\textsc{ii}] image 
to detect the fainter H$\alpha+$[N\textsc{ii}] emission.

The radio properties of 2A~0335$+$096 are also of note. Sarazin et al.
(1995) find an amorphous, steep spectrum ($\alpha \sim -1.4$) 
`mini-halo' around
PGC 013424 which has relatively low radio power compared to 
other similar systems. Fig.~\ref{fig:percentile} shows the image of the
percentile variation  ($(\Sigma_{cl} - \Sigma_{mod})/ \Sigma_{cl}$) 
of the cluster X-ray surface brightness $\Sigma_{cl}$ with respect 
to the best fit double $\beta$-model $\Sigma_{mod}$. 
The isocontours of the VLA 1.5~GHz C$+$D-array image (resolution 16\arcsec)
from Sarazin et al. (1995) are superimposed.
The image clearly shows the position of the X-ray front 
together with the X-ray blobs and the Western and Eastern holes earlier 
studied in \S\ref{par:front} and \S\ref{par:core_small}, respectively.
We notice that the mini-radio halo is well within the radius
defined by the position of the X-ray front. 
Unfortunately the spatial resolution of the radio map is insufficient
to correlate it with the X-ray blobs. However,
there is some indication that, as observed in other clusters 
(see McNamara 2002 for a review), both X-ray holes may contain 
some steep spectrum radio emission which may be `relic' emission
from previous radio lobes. Future high resolution, low frequency
radio observations will be crucial to linking the radio mini-halo and
X-ray structure.

It is worth noting that Fig.~\ref{fig:percentile} also suggests the presence
in \2A ~ ghost X-Ray bubbles at larger radii similar the ones 
observed in other clusters  (McNamara 2002).
Unfortunately, because of the relatively short exposure,
these bubbles are not statistically significant.
A deeper X-ray observation is needed to confirm their existence.

\section{Discussion}\label{par:discussion}

The \chandra ~ observation of \2A ~ reveals that  the core region
of this otherwise relaxed cluster is very dynamic. 

The cluster image shows an edge-like feature at $40-90$~kpc
from the X-ray peak to the south. The edge spans a sector from
150\degd ~ to 210\degd~ from the North.
 
On smaller scales ($<20$~kpc) the cluster image shows a very complex
structure with  X-ray filaments or blobs.  Moreover it shows
the presence of two X-ray cavities (see Fig.~\ref{fig:im_zoomed}).

The spectral analysis clearly shows that on large scales the cluster
temperature distribution is not azimuthally symmetric (see \S\ref{par:tmap}
and \S\ref{par:tprofile}).  Furthermore,  the cluster gas
temperature reaches its maximum in a region at $\approx 90$~kpc from the
X-ray peak to the south, just beyond the surface brightness edge
(Fig.~\ref{fig:temp_map}).  On smaller scales the X-ray
blobs have similar temperatures but different pressures.

Below we discuss the results of our analysis and propose 
possible dynamical models that may explain the observed 
properties of the core of \2A .

\subsection{Nature of the surface brightness edge: 
a cold front}\label{par:nature}

In \S\ref{par:front} we analyze the spatial structure of the X-ray front.
We show that the front is well fit by a discontinuous density profile with a
density jump at the front of a factor $\approx 1.6$.  Similar density jumps
have been  observed by \chandra ~ in many clusters,
including A2142, A3667, RX~J1720+26, A1795.  In those clusters, they are
interpreted as sharp boundaries of a dense, cooler gas cloud moving with a
speed $v$ through the hotter and more rarefied ambient gas.  The surface
brightness edge in \2A ~ appears to be a similar cold front. As discussed
by Vikhlinin, et al. 2002a, the pressure profile across the front can be
used to determine the Mach number $M\equiv v_1/c_{s1}$ of the moving cooler
gas cloud with respect to the speed of sound in the gas upstream, $c_{s1}$.

In  Table~2 we show that the pressure jump
across the front is $A_{P}=1.6\pm0.3$. This corresponds to a Mach 
number for the dense central gas
cloud of $M\simeq 0.75\pm0.2$ (see e.g.
Fig.~6 of Vikhlinin, et al. 2001a). Such subsonic motion of the cool gas
cores appears to be very common among the otherwise relaxed clusters with
cooling flows (Markevitch, Vikhlinin, \& Forman 2002).

\subsection{X-Ray Blobs}\label{par:dis_blobs}

In the analysis above, we identified a number of X-ray blobs most of which
are likely to be part of a filamentary structure.  As can been seen from
Table~3, the blobs are quite dense: their densities range from $\approx
0.06$--$0.13$~cm$^{-3}$.  Although the projected spectral properties of the
blobs are significantly different (i.e. they have different emission
weighted temperatures; see right panel of Fig.~\ref{fig:temp_map} and
Fig.~\ref{fig:hardness}), in \S\ref{par:cavities} we show that the
temperature of their immediately surrounding ambient gas and the deprojected
blob temperatures are consistent within the errors for all but one of the
blobs (see upper panel Fig.~\ref{fig:pressure}).  This is particularly
interesting as it implies that the thermal pressure of the blobs is higher
than the thermal pressure of the surrounding ambient gas.  This finding,
however, does not necessarily mean that the blobs are out of pressure
equilibrium with the ambient gas as the latter may have extra
non-thermal pressure (for example, magnetic fields and/or relativistic
electrons ejected from the central AGN).  Although at the moment it is not
clear which physical processes may be responsible for the formation of the
observed substructures, we can safely state that the system is dynamically
unstable and the presence of blobs is likely to be a transient phenomenon.
We notice that the evolution of the system strongly depends on the total
pressure supporting the ambient gas.  Below we discuss two extreme
possibilities:

1) the ambient gas surrounding the blobs is only supported by thermal
pressure and thus the blobs are actually not in pressure equilibrium;

2) the ambient gas, and possibly to some extent the blobs too, are supported
by an extra non thermal pressure component such that the blobs and the
ambient gas are actually in pressure equilibrium.

In the first case, the blobs are overpressured so we expect them
 to expand.  The expansion time
is of the order of the blob's  sound crossing time:
\begin{equation}
\begin{array}{rl}
t_{cross}&=2r_1{\sqrt{\mu\over \gamma kT}}= \\ 
& =1.7\times 10^{7} {\rm yr}  
\left( {r_1 \over 5 {\rm kpc}} \right)\sqrt{(\mu/0.69) m_p \over (\gamma/1.66) (kT/1.3 {\rm keV})} \, ,
\end{array}
\end{equation}
where $r_1$ is the radius of the blob, and $\mu$, $m_p$, and $\gamma$ are the 
mean particle weight, the proton mass and the adiabatic index, respectively.  
Using the linear dimension of the blobs reported in Table~3 we find 
 $t\approx 1-2 \times 10^{7}$~yr, which appears to be very short compared 
with the likely cluster age. Furthermore,  
the blobs cooling time is $\approx 10$ times 
longer than the dynamical time
 (see lower panel of Fig.~\ref{fig:pressure}).
Thus, in this first case in which
 the external gas is not 
supported by an extra pressure component, the blobs should 
expand before their excess pressure is radiated away.
Consequently we do not expect to have a significant mass deposition 
into a cold phase from the blobs.

Conversely, if the blobs  
are both in thermal and pressure equilibrium with the ambient gas, as in
the second  case considered above, the blobs are likely 
to be more stable. Nevertheless, as the blobs are heavier than the 
ambient gas (they have higher densities), 
they are expected to  sink eventually toward the minimum of the 
potential well.
The time scale $t_s$ needed by the blobs to reach the  cluster center
is obviously $t_s>> t_{ff}$\footnote{Note that $t_s= t_{ff}$ just in 
the extreme case 
in which we assume that both the  gas is collisionless and the blob 
have zero angular momentum.}, where $t_{ff}$ is the free fall time.
If we assume the cluster potential well is described by a  NFW profile   
(Navarro, Frenk \& White, 1996) the free fall time for a  point mass 
at rest at a radius  $r<<r_s$ can be well  approximated by:
\begin{equation}
t_{ff}=\sqrt{2r\over g}, \\
\end{equation}
where $r_s$, is the scale radius, $g=2 \pi G \delta_c\  \rho_{crit}$, 
$G$ is the Newton's constant, 
$\rho_{crit}=3H^2/8\pi G$ is the critical density for closure,
and $\delta_c$ is the characteristic density contrast. 
Using the best fit NFW parameters for our cluster and  $r\approx 20$~kpc 
we find:
\begin{equation}
t_{ff}\approx 108 \times h_{70}^{-1}\sqrt{r\over 20~{\rm kpc}} {\rm yr}. 
\end{equation}
From Table~3 we see that the free fall and  the cooling times are similar,
thus the actual sinking time is much longer than the cooling time. 
Therefore, unlike the first case considered above,
if the ambient gas is supported by an extra non thermal pressure component
such that the gas and the blobs are in pressure equilibrium,
we expect the blobs to cool down before they can actually  sink 
into cluster center. Hence, some mass deposition into a cold phase 
is  possible.

\subsection{Possible dynamical models}

We conclude this section by discussing possible 
dynamical models that may explain the origin of the observed
complex structure in the core of \2A .
Our starting point is the evidence  for the presence of 
a dense gas core that moves from North to South,
with a Mach number  $M\simeq 0.75\pm 0.2$ (see \S\ref{par:nature}). 

We propose two  models:\\ 
i) the motion of the cool core induces
instabilities that penetrate inside the core and disturb the gas,\\ 
ii) the central galaxy has an intermittent AGN which makes the cool 
gas ``bubbling''.\\
The features observed in the cluster result from the development of
hydrodynamic instabilities induced by these two phenomena.  Below we discuss
in detail these two models. We conclude that both models are consistent with
the available data and that further observations are required to
discriminate between the two cases.

\subsubsection {Core Destruction by Hydrodynamic Instabilities}

We first show that small-scale structure can be the result of hydrodynamic
instabilities induced by the observed motion of the core gas (revealed by
the presence of the cold front.  This motion may be due to a merger (as in
A2142 or A3667, Markevitch et al.\ 2000; Vikhlinin et al.\ 2001), or due to
the gas sloshing as in many other cooling flow clusters (Markevitch et al.\
2001, 2002). The precise nature of the core motion is not important.

The orientation of the front (see e.g. Fig.\ref{fig:percentile}), together
with the asymmetry in the temperature distribution (see
Fig.\ref{fig:temp_map}), suggests that the core is moving along a
projected direction that goes from north to south. If the core is in fact
a merging subcluster, it is clear that in projection the position of 
the subcluster is very close to the cluster center (see \S\ref{par:core}),
from the available data, however, we cannot measure the merger impact 
parameter.
Hence, the subcluster may in fact be passing through the cluster at any
distance along the line of sight from the center.

When a dense gas cloud is moving
with respect to a rarefied one,  the interface between the two gasses
is subject to hydrodynamical   Rayleigh-Taylor
(R-T) and Kelvin-Helmholtz (K-H) instabilities.  
In the following we estimate the effect of both instabilities  
on the moving core using the same
approach as  Vikhlinin \& Markevitch (2002). 
We assume that the gas cloud is a sphere
of radius equal to the distance of the observed front, $R\approx
56.7$\arcsec ~($\approx 39.7$~kpc).  As the cloud moves, the external gas flows
around its border. The  inflowing gas slows down at the leading edge of
the sphere but reaccelerates at larger angles as it is squeezed to the
sides by new portion of inflowing gas.  In the case of an incompressible
fluid, the velocity at the surface of the sphere is purely tangential
and is given by
$v_{fluid}=3/2 v_{\infty}\sin \varphi$, where $v_{\infty}$ is the velocity
of the inflowing gas at infinity and $\varphi$ is the angle of the
considered point on the cloud with respect to its direction of motion (see
e.g. Lamb 1945).  Unfortunately there is no analytic solution for the flow
of a compressible fluid around a sphere. Nevertheless, the qualitative
picture is similar. As we are only interested in order of
magnitude estimates, in the following we parametrize the actual fluid speed
introducing the parameter $\eta$ so that: 
\begin{equation}
v_{fluid}=\eta v_{\infty}\sin \varphi. 
\label{eq:v}
\end{equation}
The parameter $\eta$ should range between 1 
(fluid velocity at the surface of the sphere equal to the 
inflowing gas at infinity) and 1.5 (incompressible fluid).

First we consider the R-T instability.
This instability develops only if the acceleration  
induced by the drag force $d_g$ is greater than the  
gravitational potential acceleration near the front due to the cluster mass 
inside the moving cloud.
If we assume hydrostatic equilibrium, then 
the gravitational acceleration 
is given by:
\begin{equation}
g=-{kT\over \mu m_p R}\left ( {{d\log n \over d \log r }+{d\log T\over d \log r}} \right),
\end{equation}
where, $\mu m_p$ is the mean particle  mass, and $n$ and $T$ are the gas
electron density and temperature profiles, respectively.  
In our case this gives  $g=1.7\times 10^{-8}$~cm~s$^{-2}$ 
(see \S\ref{par:front}, 
\S\ref{par:tprofile_NS}, and Fig.~\ref{fig:cold_front}).
On the other hand, the cloud acceleration due to the drag force is given
\begin{equation}
g_d=0.15 {n_{in}\over n_{out}} {Mc_s\over R},
\end{equation}
(where $n_{in}$,$n_{out}$ the gas electron density inside and outside the 
edge) which in our case gives $g_d=2.5\times 10^{-9}$~cm~s$^{-2}$. 
As the drag acceleration is less than the gravitational one, then the 
R-T instabilities should be suppressed.

Now we consider the K-H instability. This instability may develop whenever 
two fluids in contact have a non-zero tangential velocity.
As for the R-T instability, the presence of a gravitational field 
may  suppress the development of the instability.
The stability condition is such that  
a perturbation with wavenumber less than $k=2\pi/\lambda$
is stable if:
\begin{equation}
g\ge  {Dk(v)^2\over D^2-1}
\label{eq:stab}
\end{equation}
where $g$ is the gravitational acceleration at the interface between the 
two regions and $D\equiv n_{in}/n_{out}$ is the density contrast.
Using the Eq.~\ref{eq:v} for the tangential velocity,  
Eq.~\ref{eq:stab} becomes:
\begin{equation}
\lambda\ge  2\pi {D\over D^2-1} {\eta^2 v_{\infty}^2  \sin^2\varphi \over g}
\approx 400-900  \sin^2\varphi \, {\rm kpc}.
\label{eq:lambda}
\end{equation}
Eq.~\ref{eq:lambda} shows that gravity suppresses  
instabilities only  on scales larger than $400$~kpc. Hence, 
in absence of other stabilizing mechanism, we expect the formation of
the K-H instabilities on smaller scales. 

In the case of \2A ~, we see  the instabilities 
on scales of the order of the radius of the sphere $R$ 
may develop already at  $\varphi> 18$\degd .
The condition that  the instability on the gas cloud scale can develop,
although necessary is not sufficient. In fact, in order to
effectively disrupt the gas of the moving gas cloud, the
growth time  $\tau$ of the instability must be much shorter than the 
dynamical time of the system (the 
cluster crossing time, $t_{cross}$).
We thus require (see Vikhlinin, Markevitch, \& Murray  2001b, 
and algebraic correction of 
Mazzotta et al. 2002):
\begin{equation}
{t_{cross}\over \tau}\sim 3.3 {L\over \lambda} \sin \varphi>>1.
\end{equation}
If we assume that the cluster radius is $r=1$~Mpc,
 at the angle at which the instability starts to develop
of $\varphi=18$\degd, we find that $t_{cross}/\tau\approx 25$. Furthermore, 
at the much larger angle of   $\varphi=90$\degd, 
where the instabilities are most effective, we find
$t_{cross}/\tau\approx 83$. Therefore, they 
have sufficient  time to grow to the non-linear regime. 

The effect of K-H instabilities is to turbulently mix the gas in the 
dense moving cool cloud with the more diffuse hotter cluster gas.
In particular some less dense gas from the cluster could be deposited 
into the denser gas cloud and vice-versa. Such a process may be responsible
for the formation of the observed X-ray clumpy structure.
Similarly,  any pre-existing cold ($<30$~K) gas deposited in the
cluster and/or sub-cluster core by a cooling flow will also be mixed
and re-distributed on these scales. This would explain the observed
H$\alpha+$[N\textsc{ii}] emission morphology. On the other hand, these
sub-sonic K-H instabilities will not produce any significant radio
signature. Hence, in this scenario the radio properties of \2A ~ 
(the mini-halo and the X-ray 
holes) must be related to activity of the central galaxy.

In conclusion,  the observed X-ray properties of the \2A ~ core may
be produced by  K-H instabilities induced by the motion of the core.

\subsubsection{Gas ``bubbling'' by central AGN}

The observed structure of \2A ~ can also be accounted for if the central
cluster galaxy hosts an  AGN that undergoes intervals of strong activity
followed by periods of relative quiescence (e.g. Binney \& Tabor 1995, Soker
et al. 2000).  During each active period, the AGN inflates two bubbles by
filling them with relativistic electrons.  As observed in other clusters
(e.g.  Perseus, Fabian et al. 2000; Hydra A, McNamara et al. 2000; Abell
2052, Blanton et al. 2001), this expansion creates an external shell where
the gas is denser.  The bubbles rise buoyantly, entrain the dense gas
underneath them and induce convection (e.g., Churazov et al.\ 2001).
These bubbles are also subjected to hydrodynamic instabilities that destroy
them, fragmenting  the denser shell into what we observe as denser
X-ray blobs.  The
observed structure in the core may be created by numerous bursts of AGN
activity during which the jets point in different directions.  We can
visualize this process as ``gas bubbling'' induced by the energy
injection from a central AGN.  The difference from other clusters with
similar X-ray cavities but lacking such filamentary or blobby structure
might be caused by a shorter duty cycle of the central AGN in \2A , of the
order of $10^{7}$~yr (see \S\ref{par:dis_blobs}).

The relativistic electrons produced by the AGN that are likely to be present
inside the non-thermal bubbles (see e.g. McNamara 2002 and references
therein) should eventually distribute in the central cluster region.
However, they  may avoid the densest gas blobs 
due to, for example, an enhanced magnetic field  
created during the shell compression induced by the bubble expansion.
These electrons would add a non-thermal 
pressure component in the ambient (inter-blob) gas that may
account for the apparent pressure nonequilibrium of the cluster core. 

As discussed  in \S\ref{par:dis_blobs}, in this case the dense X-ray
blobs have time to cool, depositing some mass in a cold phase.  This
cool gas may explain the presence of the observed CO and 
H$\alpha+$[N\textsc{ii}]
extended emission in the cluster center (see Fig.~\ref{fig:halpha}).

The relativistic electrons produced by the AGN may also account for the
presence of the mini radio halo observed in the cluster core (see
Fig.~\ref{fig:percentile}).

We finally mention that there may be a connection between the AGN activity
and the core sloshing, hinted at by the observed correlation between the
presence of the X-ray bubbles and cold fronts in the central regions of many
cooling flow clusters (Markevitch et al.\ 2002).  Reynolds, Heinz \&
Begelman (2002) showed that during an AGN jet formation, a spherical pulse
forms and propagates at the sound speed into the intracluster medium.
Repeated AGN activity may inject sufficient kinetic energy to produce
sloshing of the gas, although it is unclear how the bulk motion of the core
as a whole could be produced.  Sloshing of the dense core may also be
induced by a merger, even if the merging subcluster did not reach the center
of the main cluster (e.g., Churazov et al.\ 2003).

To conclude this section, ``gas bubbling'' induced by central AGN 
provides a plausible explanation of the observed features 
in the cluster core.

\section{Conclusion}\label{par:conclusion}

The poor cluster \2A ~ has several remarkable features that are only
detectable with the superb resolution of \chandra . First, there is a cold
front indicating a mildly sub-sonic gas motion. Second, two X-ray cavities
that may be associated with steep spectrum radio emission indicating
previous radio activity in this system. Third,  a number of small dense
gas blobs in the cluster core that may be the shreds of a cooling core
disturbed either by K-H instabilities or ``bubbling'' induced by
intermittent AGN activity.  All of these properties relate to processes
that may act to  disrupt or destroy any cooling flow (sub-cluster
merger, injection of non-thermal electrons into the intracluster medium
 or direct radiation
from an AGN). From this relatively short X-ray observation it is not
possible to determine which process (if any) dominates. Future deeper
observations in the X-ray, radio and optical of \2A ~ may allow the
energetics of each to be assessed. When combined with observations of other
similar clusters, it will be possible draw concrete conclusions about the
astrophysical nature of cooling flows.

\acknowledgments
 
We thank David Gilbank for useful comments and suggestions and for assistance
with the narrow-band optical image, Craig Sarazin for providing the VLA
images, and Alexey Vikhlinin for helpful discussions. This work was
supported by an RTN fellowship, CXC grants GO2-3177X and GO2-3164X, the
Royal Society and the Smithsonian Institution.

\end{document}